\documentclass[12pt]{article}
\usepackage{graphicx,epsfig,epstopdf}
\usepackage{color}
\usepackage{caption}
\usepackage{eqnarray,amsmath}

\def\RL{R_{L}}
\def\Cp{C_{p}}
\def\VCC{V_{CC}}
\def\VEE{V_{EE}}
\def\Rf{R_{f}}
\def\Pr{P_{r}}
\def\Vo{V_o}

\title{Pulsed Self-Oscillating Nonlinear Systems for Robust Wireless Power Transfer}
\author
{Fu Liu,$^{\ast}$ Bhakti Chowkwale, Prasad Jayathurathnage, Sergei Tretyakov\\
	\\
	\normalsize{Department of Electronics and Nanoengineering, Aalto University,}\\
	\normalsize{P.~O.~Box~15500, FI-00076 Aalto, Finland}\\
	\normalsize{$^\ast$E-mail: fu.liu@aalto.fi}
}
\date{}

\begin{document}


%

\maketitle

\begin{abstract}
While wired-power-transfer devices ensure robust power delivery even if the receiver position or load impedance changes, achieving the robustness of wireless power transfer (WPT) is challenging. Conventional solutions are based on additional control circuits for dynamic tuning. Here, we propose a robust WPT system in which no additional tuning circuitry is required for robust operation. This is achieved by our systematically designing the load and the coupling link to be parts of the feedback circuit. Therefore, the WPT operation is automatically adjusted to the optimal working condition under a wide range of load and receiver positions. In addition, pulsed oscillations instead of single-harmonic oscillation are adopted to increase the overall efficiency. An example system is designed with the use of a capacitive coupling link. It realizes a virtual, nearly-ideal oscillating voltage source at the load site, giving efficient power transfer comparable to that of the ideal wired-connection scenario. We numerically and experimentally verify the robustness of the WPT system under the variations of load and coupling, where coupling is changing by our varying the alignment of aluminum plates. The working frequency and the transferred power agree well with analytical models. The proposed paradigm can have a significant impact on future high-performance WPT devices. The designed system can also work as a smart table supporting multiple receivers with robust and efficient operation.
\end{abstract}

\clearpage
\section{Introduction}\label{sec:intro}
The concept of wireless power transfer (WPT) was patented by Tesla about 100 years ago \cite{tesla1914}. Ever since, researchers and engineers have been investigating methods for improving WPT technologies in terms of efficiency, power, transmission distance, human safety, and adaptive tuning \cite{Hui2014,Kurs2007,Sample2011,Christ2012,Song2017}. Recently, WPT technologies have been widely used in many applications including consumer electronics \cite{Chao2000}, biomedical implants \cite{Kakegawa1981,Ho2014}, industrial applications, and electric vehicles \cite{Vilathgamuwa2014,Dai2015a}. A typical WPT system consists of a high-frequency ac generator, a transmitter, and a receiver that is separated from transmitter and connected to the electrical load \cite{Hui2014}. Energy is transferred from the transmitter to the receiver via a wireless link using either inductive coupling (magnetic field) \cite{Hui2014,Kurs2007,Sample2011} or capacitive coupling (electric field) \cite{Dai2015a,Kline2011,Dai2015,Lu2018}. The optimal power transfer occurs when the frequency and load are chosen to ensure the conjugate match of the source and load impedances. However, in practical applications, variations in working conditions such as the receiver position (coupling strength) and load resistance are inevitable, and cause the system to deviate from optimal operation \cite{Sample2011}. Thus, a WPT system optimized for a nominal working condition may not function optimally when the working conditions change. Conventionally, this challenge is addressed by adding control circuits to sense the operating state and tune the system to maintain optimal conditions~\cite{Sample2011,Beh2013,Duong2015,LiuKeshuang2018,Xu2017}, but such circuits add to the cost and complexity of the system. Therefore, a radically new method to achieve the system robustness without any additional tuning under system variations is highly demanded for WPT devices.

Recently, a parity-time symmetric system consisting of coupled resonant circuits with gain and loss was shown to enable wireless power transfer that is robust to variations in coupling through self-tuning of the system's oscillation frequency~\cite{Schindler2011PTsymmetricLRC,Assawaworrarit2017,Radi2018}. However, the implementation of a gain circuit relied on negative-impedance converters, which exhibit significant losses during low dc-to-ac power conversion~\cite{Assawaworrarit2017,Pozar2011}. This results in a low end-to-end transfer efficiency when the total power supplied to the generator is accounted for, including the dc power dissipated by the amplifiers. A more recent study of on-site WPT showed that robust transfer can be achieved with a generic amplifier to generate oscillations at the load directly~\cite{Radi2018}. The realization reported in Re.~\cite{Radi2018} uses resonant time-harmonic oscillations, which are generated through interactions between the amplifier and the load. This resonant approach, however, results in unavoidable inherent losses as it either requires consumption of dc power to bias the amplifier in the linear region, or results in conversion of energy into higher-order harmonics of the resonant frequency. The design of a robust yet nonresonant WPT system that can efficiently transfer pulsed signals (e.g. square wave) could enable increased overall efficiency for practical nonlinear amplifiers. In this work, we propose and realize a robust and efficient on-site WPT system working in the nonresonant pulsed regime. While the self-oscillating comparator-based relaxation oscillator has been known for nearly a century~\cite{Abraham1919,Ginoux2012}, the WPT system proposed here is a modified one where the wireless link and the load are designed to be parts of the oscillator's feedback circuit. This topology provides the advantage of detecting the receiver and automatic initiation of wireless power transfer. This is different from the earlier solutions which use self-oscillating circuits to enhance robustness~\cite{zierhofer1992class,ziaie2001self,ahn2013wireless}, where feedback is present inside the generator and oscillations are sustained even without the receiver. Moreover, on one hand, the self-oscillating nature ensures system robustness (i.e., the WPT system automatically adjusts itself to the optimal working condition although the load resistance and the receiver position vary over a wide range). On the other hand, the pulsed oscillations from the switched circuits working in a highly nonlinear regime guarantee that, within a certain range of  operation conditions, the proposed system can achieve high end-to-end power-transfer efficiencies, approaching the efficiency of the ideal case in which the load is directly connected to an ideal voltage source with zero internal resistance. The robust and efficient operations are demonstrated numerically and experimentally with a capacitively coupled WPT system.

\section{The pulsed self-oscillating WPT system}\label{sec:compWPT}

\begin{figure}[t!]
	\centering
	\includegraphics[width=0.8\columnwidth]{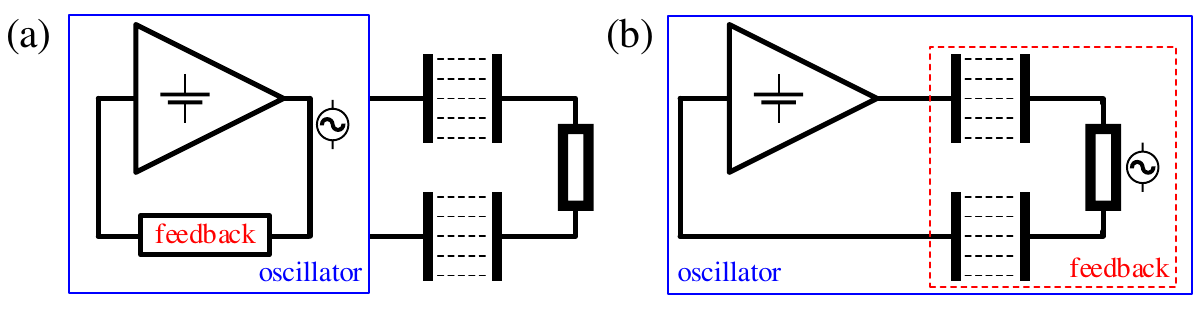}
	\caption{\label{fig1-comp} Comparison between the conventional and proposed WPT systems. (a) In conventional systems, the high-frequency ac oscillation is formed in the oscillator (generator) and then transmitted to the load through a capacitive wireless link. (b) In the proposed WPT system, the wireless link and the load are parts of the feedback circuit. The whole system forms a unified oscillator and the oscillating power is directly generated at the load.}
\end{figure}

The conventional WPT system and the self-oscillating WPT system are compared in Fig.~\ref{fig1-comp}, with capacitive WPT as an example. In conventional devices, the generator (oscillator), transmitter and receiver are functionally distinct. The utility power is first converted to high-frequency ac power by use of an oscillator since dc or 50-60-Hz power cannot be effectively transferred via free space. Next, the output power is delivered from the transmitter to receiver through capacitive coupling elements. The operating conditions of the generator are determined by its circuitry, therefore, the transferred power changes with the variation of the load and the receiver positions. Conversely, in the proposed self-oscillating WPT paradigm, the load and the wireless power link are parts of the positive-feedback loop, forming a unified oscillator circuit. The high-frequency ac power is therefore directly generated at the load position, as shown in Fig.~\ref{fig1-comp}(b). The feedback loop passing through the coupling capacitance and the load determines the oscillation frequency by itself. In this way, feedback of the distributed oscillator maintains the optimal condition without any additional tuning circuitry.

\begin{figure}[t!]
	\centering 
	\includegraphics[width=0.8\columnwidth]{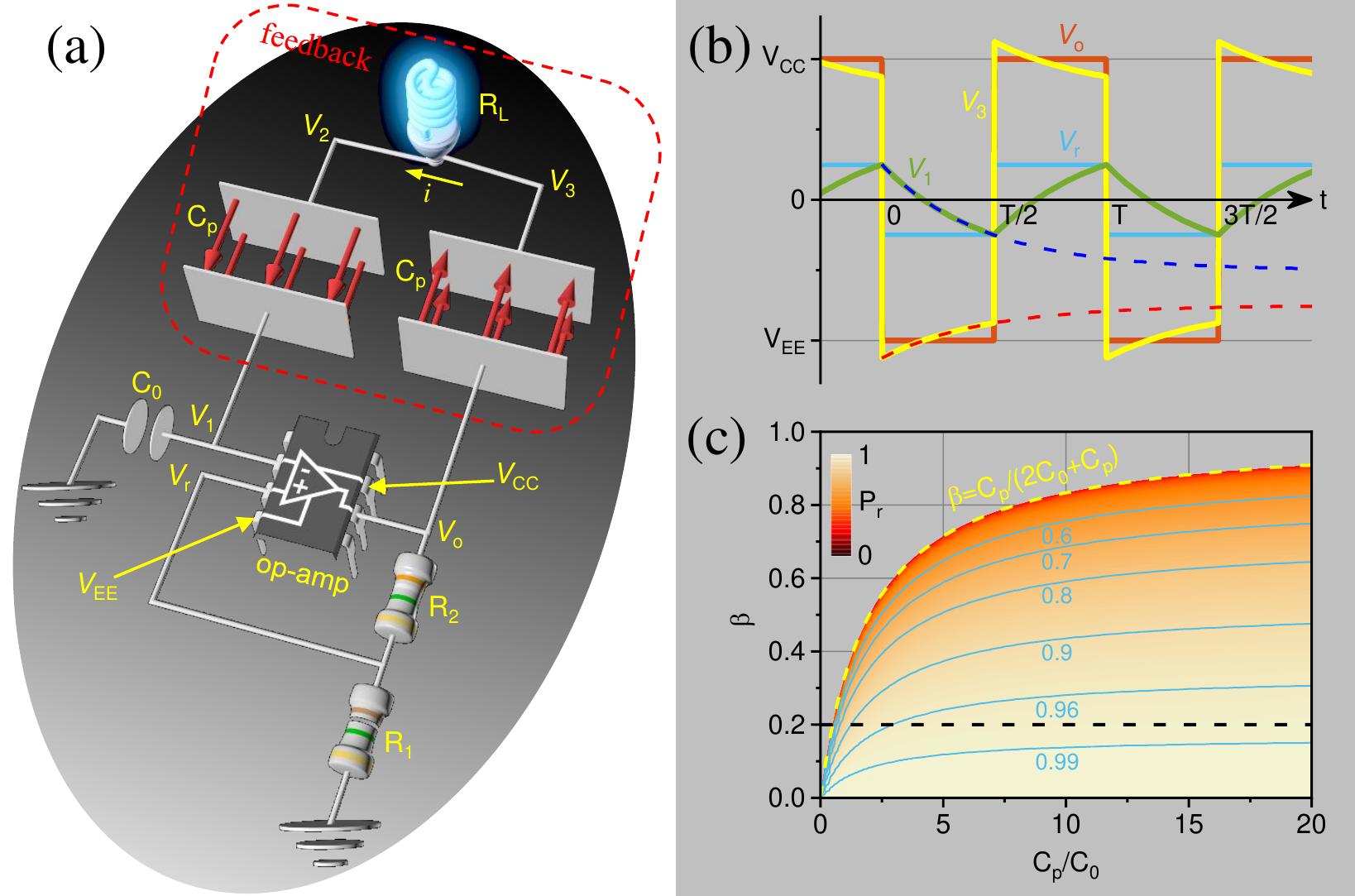}
	\caption{\label{fig2-schem} The proposed WPT system based on a self-oscillating circuit and its theoretical performance. (a) The WPT system with variable definitions. The feedback loops at inverting (-) and noninverting (+) terminals of the operational amplifier make the output oscillate, driving the energy wirelessly to the load. (b) Voltage oscillations from simulations (solid lines) and analytical solutions of $V_1$ and $V_3$ in Eqs.~(\ref{Eq:solV1}) and~(\ref{Eq:solV3}) (dashed lines). (c) Analytical power ratio $\Pr$ as a function of $\Cp/C_0$ and $\beta$. The dashed yellow line indicates the boundary of the oscillation regime. The system operates closer to the ideal wired-connection case when the configuration is further away from the boundary. The horizontal dashed black line is for $\beta=0.2$.}
\end{figure}

As a first conceptual test to demonstrate the proposed principle, we suggest a self-adaptive WPT system in the pulsed regime [see Fig.~\ref{fig2-schem}(a)]. In this system, an operational amplifier \cite{Graeme1973} is used to work as a switch to generate  pulse oscillations. The two identical parallel-plate capacitors $\Cp$ provide a wireless link for transferring energy to the load $\RL$, as well as a positive feedback loop to maintain self-oscillations. Example oscillations from simulations (LTspice IV, same for the following simulations) are shown as solid lines in Fig.~\ref{fig2-schem}(b). In such a configuration, if the receiver is absent, the feedback loop is disconnected, therefore there is no oscillation. However, when the receiver is brought back and the system satisfies the oscillation condition, the WPT operation will be restored. This property brings a significant advantage of automatic receiver detection which greatly reduces standby losses.

\subsection{Analytical solution of the pulsed self-oscillating WPT system}\label{subsec:analytic}

The analytical solution for the proposed WPT system in Fig.~\ref{fig2-schem}(a) is derived on the basis of the assumption that the operational amplifier is ideal. When the system operates, the same current $i$ is passing through $C_0$, $\Cp$, and $\RL$, from Kirchhoff's current law. Therefore, we obtain two voltage relations; namely, $\Vo-V_3=V_2-V_1$ and $V_2=(\Cp+C_0)V_1/\Cp$. We also obtain the master equations of the WPT system
\begin{eqnarray}
	C_0 \frac{d V_1}{d t}
		& = & \frac{V_3-V_2}{\RL},\label{Eq:master1}\\
	\Cp \frac{d (\Vo-V_3)}{d t} 
		& = & \frac{V_3-V_2}{\RL}.\label{Eq:master2}
\end{eqnarray}
Together with the initial conditions when the output voltage $\Vo$ switches from positive to negative supply voltage ($\VCC$ to $\VEE$), we obtain time-dependent voltage solutions (see Section~A in Supplemental Material~\cite{supplementalmaterial} for details) as
\begin{eqnarray}
	V_1(t) & = & \frac{\VCC}{\gamma}
		\left[(1+\beta\gamma)e^{-t/\tau}-1\right],\label{Eq:solV1}\\
	V_3(t) & = & -\frac{C_0}{\Cp}V_1(t)-\VCC,\label{Eq:solV3}
\end{eqnarray}
for the first half period ($0\leq t<T/2$), where $\beta=R_1/(R_1+R_2)$ is the voltage division factor, $\gamma=1+2C_0/\Cp$ is a dimensionless factor indicating the coupling strength, $\tau=C_0 \RL/\gamma$ is the characteristic time, and we have adopted $\VEE=-\VCC$ for symmetric operation. For the second half period ($T/2\leq t<T$), the solution can be found by simply changing $\VCC$ to $\VEE$ and $t$ to $t-T/2$. Both $V_1$ and $V_3$ change exponentially, following the charge and discharge processes of the capacitors, and they are plotted in Fig.~\ref{fig2-schem}(b) as dashed curves (blue for $V_1$ and red for $V_3$), complementing the simulation results. The oscillation period $T$ is then obtained as
\begin{equation}\label{Eq:T}
	T=2\tau\ln\frac{1+\beta\gamma}{1-\beta\gamma},
\end{equation}
through the switching condition $V_1(T/2)=\beta \VEE$. However, the condition $V_1(t\rightarrow\infty)<\beta \VEE$ must be satisfied to support oscillations. This gives the oscillation condition
\begin{equation}\label{Eq:OsciCond}
	\beta<1/\gamma.
\end{equation}
As long as the system parameters satisfy this condition, power will be wirelessly delivered to the load, meaning that the proposed WPT system is robust.

The average power delivered to the load can be calculated by $1/T\int_{0}^{T} (V_3-V_2)^2/\RL d t$, which results in
\begin{equation}\label{Eq:Pavg}
	P_{\mathrm{avg}} = \frac{\VCC^2}{\RL}\times
		2\beta\gamma\bigg/\ln\frac{1+\beta\gamma}{1-\beta\gamma}.
\end{equation}
We notice that the first term $P_0=\VCC^2/\RL$ is the power consumed by the load if we were to connect the load directly to the ideal voltage source $\VCC$ with wires having zero impedance. Thus, the value of $P_0$ provides a very good performance reference and we define a power ratio
\begin{equation}\label{Eq:Pr}
	\Pr=\frac{P_{\mathrm{avg}}}{P_0}=2\beta\gamma\bigg/\ln\frac{1+\beta\gamma}{1-\beta\gamma},
\end{equation}
to measure how well the WPT system performs compared with the direct wired connection. 

\subsection{Robustness of the pulsed self-oscillating WPT system}\label{subsec:robust}
From Eq.~(\ref{Eq:Pr}), we highlight that $\Pr$ is not a function of the load resistance $\RL$, meaning that the power ratio is robust against the changes of the load. Figure~\ref{fig2-schem}(c) shows the analytical contour plot of $\Pr$ as a function of $\Cp/C_0$ and $\beta$. As we can see, the power ratio $\Pr$ is bounded by the oscillation condition in Eq.~(\ref{Eq:OsciCond}). When the system parameters $(\Cp/C_0,\beta)$ move away from the boundary, $\Pr$ increases and can reach 0.99 for small $\beta$. High values of $\Pr$ mean that the performance of the proposed WPT system is comparable to that of the wired connection. More importantly, with constant $\beta$, $\Pr$ is insensitive for large $\Cp/C_0$, indicating that we can obtain efficient and robust WPT with high $\Pr$ for a broad range of plate capacitances (i.e., for varying coupling strength such as the transfer distance). For example, if we choose $\beta=0.2$, we can achieve $\Pr>0.9$ when $\Cp/C_0>1.23$.

\begin{figure}[t!]
	\centering
	\includegraphics[width=0.45\columnwidth]{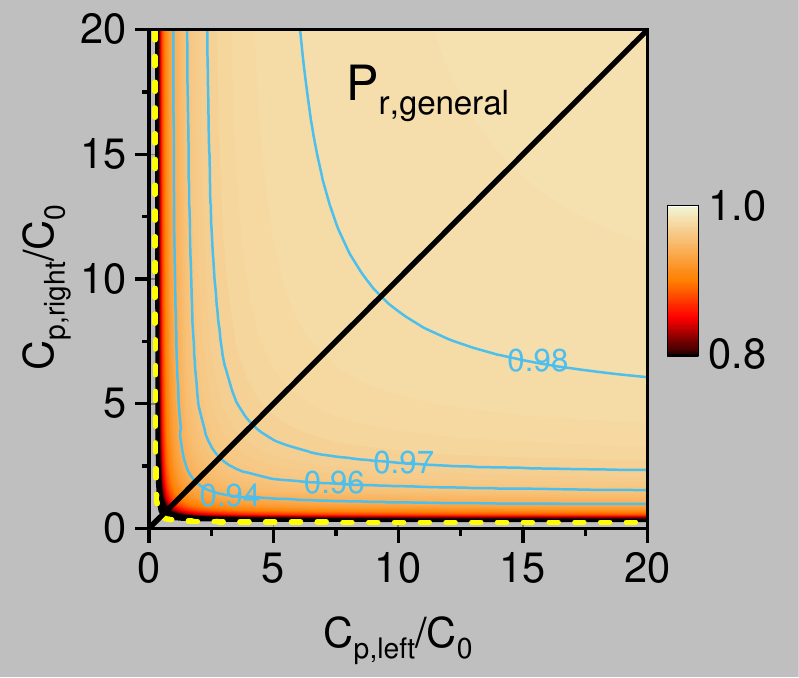}
	\caption{\label{fig3-Prgeneral} Analytical plot of the generalized power ratio as a function of $C_{\mathrm{p,left}}/C_0$ and $C_{\mathrm{p,right}}/C_0$ when $\beta=0.2$. The dashed yellow line indicates the boundary of the oscillation; that is, $1/\gamma=0.2$ from Eq.~(\ref{Eq:OsciCond}). The solid black line shows the position where the two plate capacitors have the same capacitance.}
\end{figure}

Actually, the simplifying assumption of identical plate capacitors is not necessary, and the system is still robust when they have different capacitances. For this general case, the master equation~(\ref{Eq:master2}) is slightly modified and similar results and discussions follow (see Supplemental Material for details). The generalized oscillation period, oscillation condition, delivered power, and power ratio still have the same form as Eqs.~(\ref{Eq:T})$-$(\ref{Eq:Pr}), but with generalized characteristic time $\tau'=C_0 \RL/\gamma'$ and generalized dimensionless factor $\gamma'=1+C_0/C_{\mathrm{p,left}}+C_0/C_{\mathrm{p,right}}$, where $C_{\mathrm{p,left}}$ ($C_{\mathrm{p,right}}$) is the coupling capacitance connected to the inverting (output) terminal of the operational amplifier. The slightly modified dimensionless factor $\gamma$ indicates that the asymmetric coupling capacitances from misalignments have a negligible effect on the system operation. Figure~\ref{fig3-Prgeneral} shows the analytically calculated $P_{\mathrm{r,general}}$ as a function of  $C_{\mathrm{p,left}}/C_0$ and $C_{\mathrm{p,right}}/C_0$ when $\beta=0.2$. As we can observe, when the two plate capacitances deviate from each other (being still larger than $C_0$), the change of the power ratio is relatively small, indicating that the proposed WPT system is robust against misalignments. This is of great advantage in WPT applications as misalignment of the receiving plates will not significantly affect the WPT operation.

\subsection{Comparison with the conventional WPT system}\label{sebsec:simulation}

\begin{figure}[t!]
	\centering
	\includegraphics[width=0.8\columnwidth]{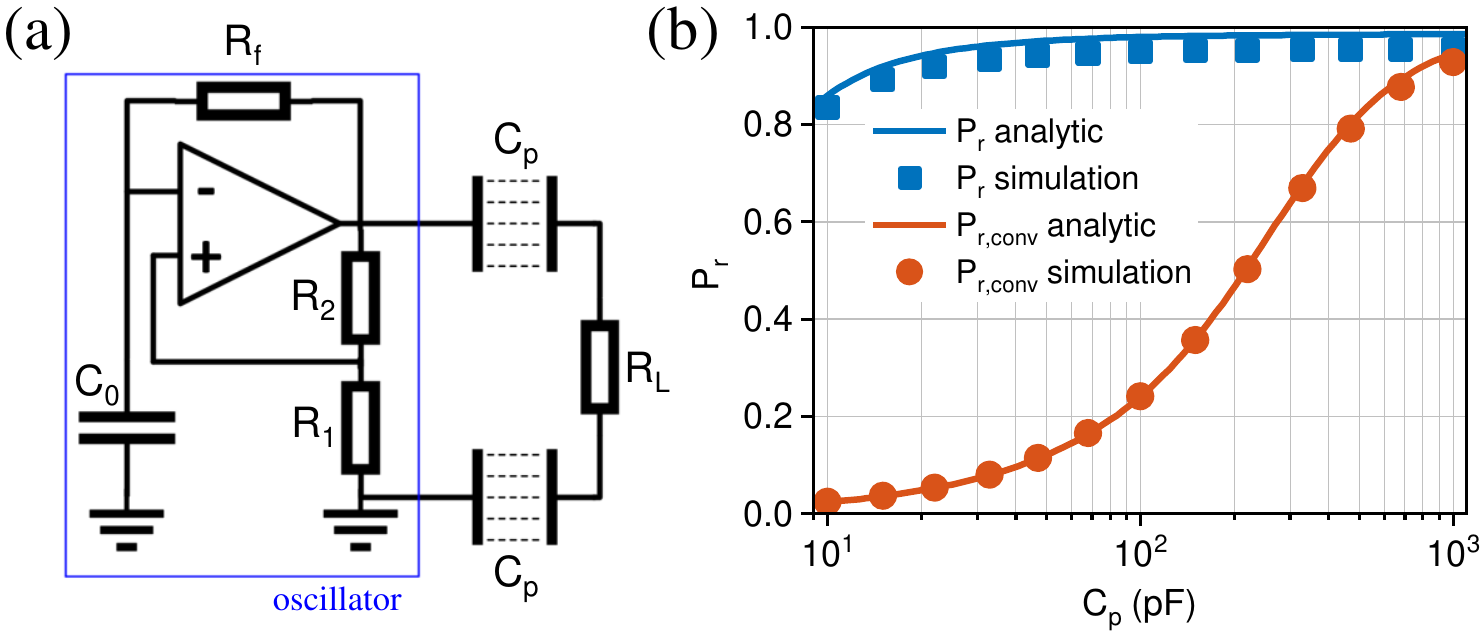}
	\caption{\label{fig4-convComp} (a) The conventional WPT system for comparison. (b) Comparison of the two power ratios against coupling $\Cp$ for the same system parameters. Simulation results are from the circuits made of nearly ideal operational amplifier.}
\end{figure}

We compare the robust operation of the proposed pulsed self-oscillating WPT system with the operation of a conventional WPT system, which is schematically shown in Fig.~\ref{fig4-convComp}(a)~\cite{Sibakoti2011}. This conventional WPT system serves as a good basis for comparison because the difference between it and the proposed WPT system is only in the location of the wireless link and the receiver, just as the difference shown in Fig.~\ref{fig1-comp}. For such a conventional WPT system, the power ratio similarly defined as Eq.~(\ref{Eq:Pr}) is found to be
\begin{equation}
	P_{\mathrm{r,conv}} = \frac{2\Cp\RL}{T_{\mathrm{conv}}} \tanh\frac{T_{\mathrm{conv}}}{2\Cp\RL},
\end{equation}
where $T_{\mathrm{conv}}=2C_0 \Rf \ln[(1+\beta)/(1-\beta)]$ is the oscillation period of the generator (see Supplemental Material for details). This power ratio is a function of the load $\RL$, indicating less robustness against load variations, compared with the load-independent one for the proposed WPT system. Moreover, in Fig.~\ref{fig4-convComp}(b) we plot the two power ratios against the change of the coupling capacitance $\Cp$ when we use the same system parameters $R_1=1~\mathrm{M\Omega}$, $R_2=3.9~\mathrm{M\Omega}$, $C_0=10~\mathrm{pF}$, and $\RL=1~\mathrm{k\Omega}$ in both WPT systems and $\Rf=100~\mathrm{k\Omega}$ for the conventional WPT system. It is clear that the proposed pulsed self-oscillating WPT system provides better robust operation than the conventional one. The power delivery of the proposed self-oscillating WPT system is comparable to that of the wired connection for coupling larger than 20~pF, which, for example, corresponds to a transfer distance of 8.8~cm for underwater WPT when the capacitors are realized on the basis of metal plates with area $5\times5~\mathrm{cm}^2$. In our particular experiment, the transfer distance is limited by the oscillation condition in Eq.~\ref{Eq:OsciCond} as a larger transfer distance (corresponds to small $\Cp$) may fail the oscillation. The low power ratio of the conventional system is due to the fixed oscillation period of the generator. When $\Cp$ is small, capacitors can be easily charged. Therefore, the WPT operation lasts for only a very short time in each half period. Indeed, this power ratio can be increased by decreasing the feedback resistance $\Rf$. However, the decrease of $\Rf$ will increase the parasitic loss (which does not exist in the proposed WPT system) and drastically diminish the maximum overall efficiency of the conventional WPT system.

We emphasize that, in principle, the proposed pulsed self-oscillating WPT system can work for any value of the load resistance and plate capacitance as long as the oscillation condition~(\ref{Eq:OsciCond}) is satisfied. In actual implementations, this means that energy can be transferred to any load with any distance and misalignment between the metal plates within the boundaries set by Eq.~(\ref{Eq:OsciCond}).

\section{Experimental demonstration of the power-transfer robustness}\label{sec:exp}

\begin{figure}[t!]
	\centering 
	\includegraphics[width=0.8\columnwidth]{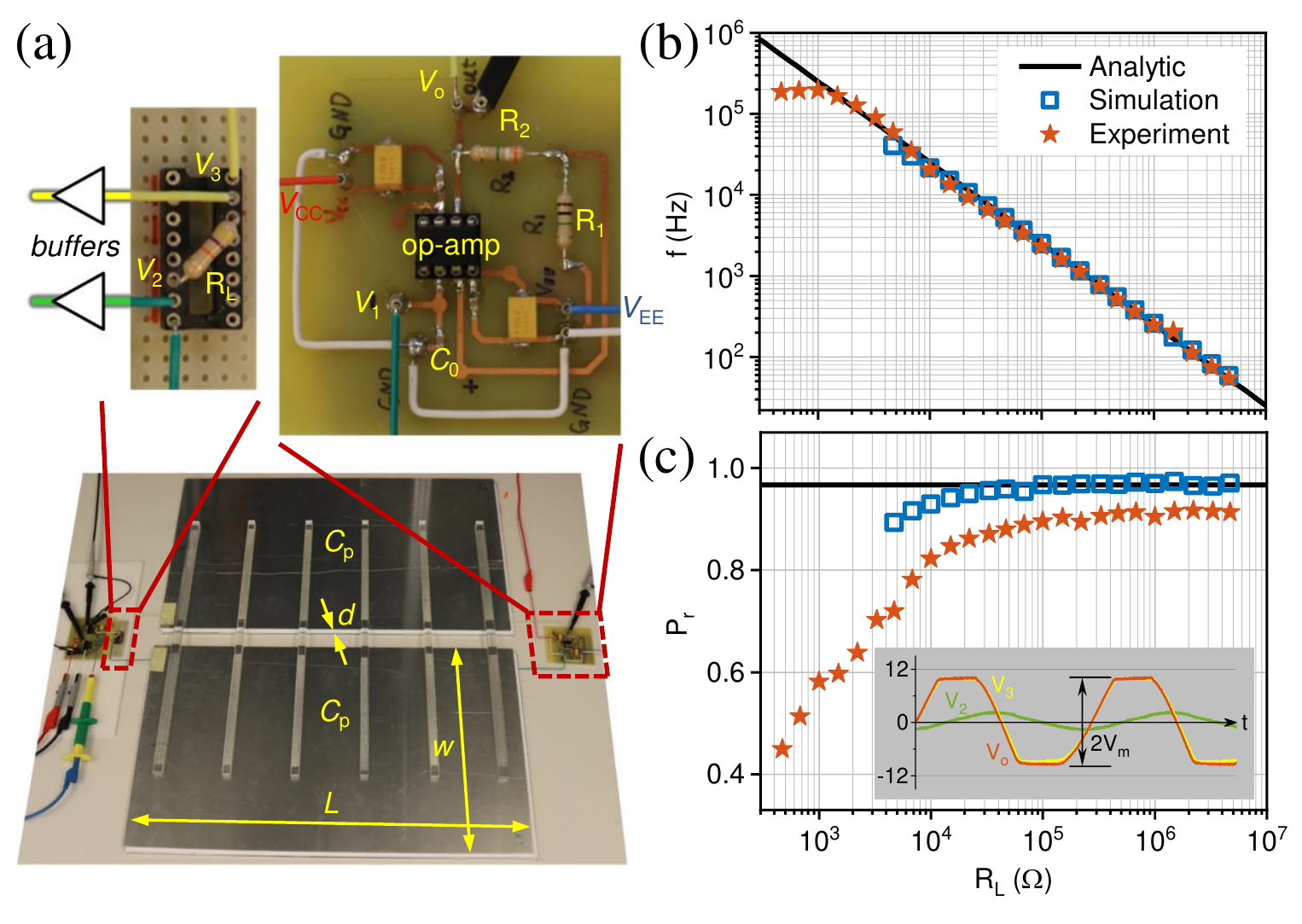}
	\caption{\label{fig5-exp4RL} Experimental demonstration of the robustness against load variation. (a) Experimental setup. The capacitive wireless links are made of aluminum plates. (b) The working frequency is inversely proportional to the load resistance. (c) Power ratio versus load resistance. The measured value drops when the load resistance decreases. The inset shows the experimental voltage oscillations when $\RL=1.5~\mathrm{k\Omega}$.}
\end{figure} 

We experimentally demonstrate the robustness of the proposed WPT system with a specific operational amplifier (TL051CP). The experimental setup is shown in Fig.~\ref{fig5-exp4RL}(a) with the system parameters $R_1=1~\mathrm{M\Omega}$, $R_2=3.9~\mathrm{M\Omega}$, $C_0=4.7~\mathrm{nF}$, and $\VCC=-\VEE=12V$. The wireless link is realized with two pairs of aluminum plates with width $w=30$~cm and length $L=50$~cm. Each pair of plates is separated by distance $d=0.1$~mm (in experiments, it is fixed with paper sheets with relative permittivity $\epsilon_r=1.4$) to provide the plate capacitance in the $\mathrm{nF}$ range, which is appropriate for the parameters of the selected operational amplifier. The transfer distance can be increased by use of high-speed switching devices instead of the general-purpose operational amplifiers used in this experimental demonstration. The three characteristic voltages $\Vo, V_2$, and $V_3$ are measured with a high-speed oscilloscope \cite{buffer}; then the data are processed to determine the oscillation frequency $f=1/T$ and the power ratio. The robustness of the WPT system is demonstrated by our varying both the load resistance $\RL$ and the receiver position (the plate capacitance $\Cp$).

\subsection{Robustness against load resistance}\label{subsec:exp4RL}

We first demonstrate the robustness with respect to changes of the load resistance $\RL$. The capacitance of the aluminum plates is measured to be $18.1~\mathrm{nF}$. In theory, this configuration should work for arbitrary $\RL$ values. In practice, the lower bound of $\RL$ is defined by the maximum oscillating frequency supported by the selected operational amplifier, which is approximately 400~$\mathrm{\Omega}$ for our device. Therefore, we vary $\RL$ from $470~\mathrm{\Omega}$ to $4.7~\mathrm{M\Omega}$, and the measured frequencies are shown in Fig.~\ref{fig5-exp4RL}(b). It is clear that the experimental and simulation results \cite{simulationlimit} agree very well with each other and the frequency is inversely proportional to $\RL$, following the analytical black line given by Eq.~(\ref{Eq:T}). In calculating the power ratio, we note that the magnitude of the $\Vo$ swing is less than the feeding voltages due to the internal voltage drop within the selected operational amplifier, as shown in the inset in Fig.~\ref{fig5-exp4RL}(c). Therefore, to give results comparable to the analytical solutions, we calculate the power ratio by $\Pr'=P_{\mathrm{avg}}/P_0'$ where $P_0'=V_{\mathrm{m}}^2/\RL$ and $V_{\mathrm{m}}$ is half of the measured peak-to-peak voltage of the output, representing a dc voltage source with magnitude $V_{\mathrm{m}}$. The corresponding power ratios from experiments and simulations are shown in Fig.~\ref{fig5-exp4RL}(c). They agree well with the analytical solution with an exception for low $\RL$ values. The system works robustly for load variations of three orders of magnitude. The load-dependent $\Pr$ from measurements comes from the finite slew rate of the selected operational amplifier, which prohibits instant switching of the output voltage $\Vo$; see the inset in Fig.~\ref{fig5-exp4RL}(c). The resulting trapezoidal shape of the output affects the shape of $V_3$ and reduces the power delivered to the load. While this power reduction is negligible for low frequencies (corresponding to large load resistance), it becomes noticeable when the oscillating frequency increases (load resistance decreases).

\begin{figure}[t!]
	\centering
	\includegraphics[width=0.45\columnwidth]{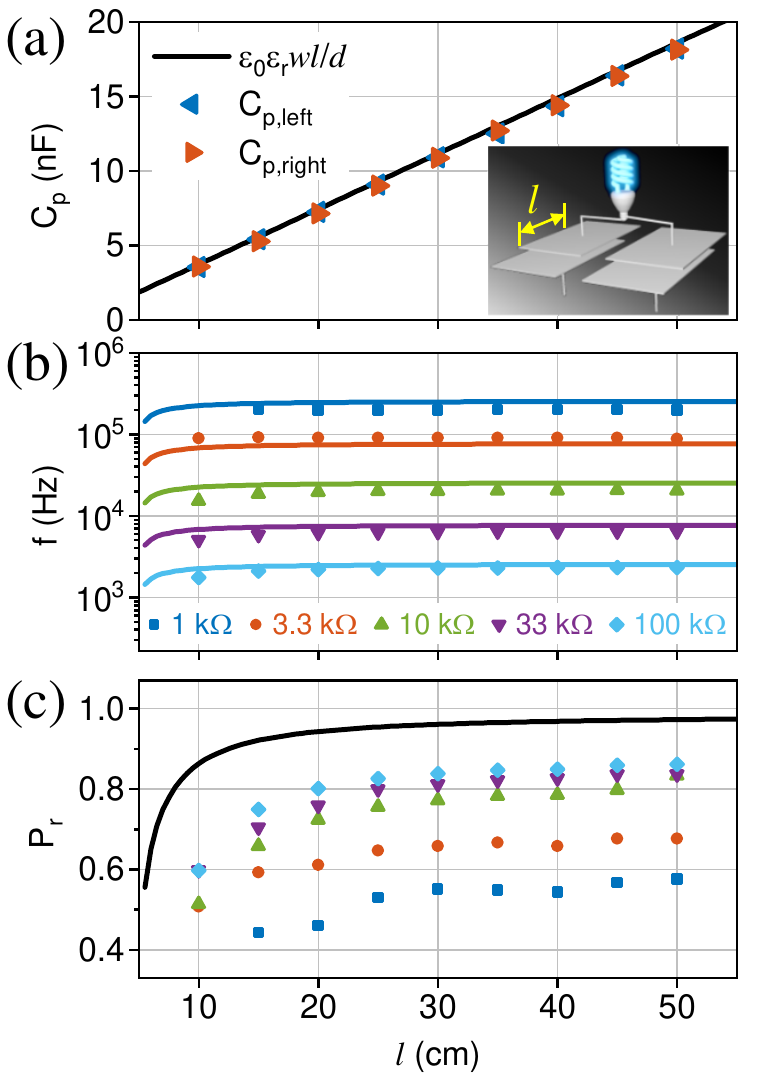}
	\caption{\label{fig6-exp4Cp} Experimental demonstration of the robustness against receiver position (overlap distance). (a) The plate capacitance is linearly proportional to the overlap distance $l$ of the aluminum plates. The inset shows how the plates are moved. (b, c) Measured (symbols) and analytical (lines) working frequency and power ratio versus the overlap distance. Five load values are measured.}
\end{figure}

\subsection{Robustness against receiver positions}\label{subsec:exp4Cp}

We next demonstrate the robustness with respect to changes of the receiver position. We modify the overlap area by sliding the top plates, as shown in the inset in Fig.~\ref{fig6-exp4Cp}(a). The plate capacitance $\Cp$ is therefore a linear function of the overlap distance $l$; the measured capacitances are shown in Fig.~\ref{fig6-exp4Cp}(a) when we change $l$ from 10 to 50~cm. We perform the experiments with five load values (i.e., $1, 3.3, 10, 33$, and $100~\mathrm{k\Omega}$). The measured frequencies and power ratios are shown in Figs.~\ref{fig6-exp4Cp}(b) and (c) as symbols, while the lines are obtained from the analytical solutions Eqs.~(\ref{Eq:T}) and (\ref{Eq:Pr}). As we can observe, the WPT system operates robustly while the receiver position is changing. For a fixed load, the working frequency is almost constant, while the power ratio increases with increase of $l$. The measured power ratio $P_\mathrm{r}$ is lower than the analytical expectations, especially for smaller load resistance $1~\mathrm{k\Omega}$ and $3.3~\mathrm{k\Omega}$. This is again due to the finite slew rate of the selected operational amplifier, the same as discussed in the previous subsection. These results are consistent with the aforementioned ones in Fig.~\ref{fig5-exp4RL}. 

Moreover, the robustness of the WPT system is also experimentally demonstrated by our moving and rotating the receiving part while the WPT system is on (see supplemental material for details). For visualization, two LEDs, accompanying the load, are glowing while the position of the receiving part changes. It is also noted that this WPT system can detect the receiver by itself and automatically start up. When the receiver is absent, the feedback loop is open and there is no oscillation, and thus no  power generation and transfer. However, when the receiver returns, the oscillations restart automatically and power is wirelessly transferred to the load, indicating by reglowing of LEDs (see the supplemental videos).

\subsection{Efficiency of the proposed WPT system}\label{subsec:efficiency}
\begin{figure}[t!]
	\centering
	\includegraphics[width=0.45\columnwidth]{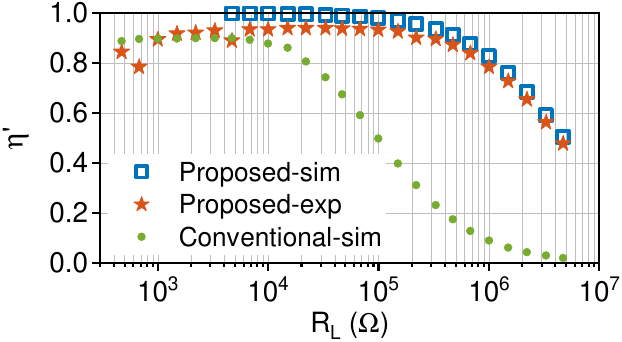}
	\caption{\label{fig7-expeff} Efficiency of the proposed WPT system (both simulation and experiment results) and the conventional WPT system (simulation results) without counting the loss in the selected operational amplifier.}
\end{figure}

In the proposed WPT system, there is no inevitable loss in the output resistance of the generator (such as $\Rf$ in the conventional WPT system shown in Fig.~\ref{fig4-convComp}) as this output resistance is now the useful load itself. In our example realization, the dissipation losses in $R_1$ and $R_2$ are negligible, and parasitic losses occur only in the wireless link and the operational amplifier acting as a switch, which are also present in all conventional systems. If we observe the overall efficiency, which is defined as the ratio between the power consumption in the load and the total power taken from the utility source, i.e., $\eta=P_{\mathrm{avg}}/P_{\mathrm{in}}$, we obtain a maximum overall efficiency of 36.7\% in our demonstration prototype. The moderate overall efficiency is due to the comparably large parasitic losses in the selected operational amplifier. On the other hand, we can calculate the efficiency without counting the loss in the operational amplifier; that is, $\eta'=P_{\mathrm{avg}}/(P_{\mathrm{avg}}+2P_{C_\mathrm{p}}+P_{R_1}+P_{R_2})$, where $P_{C_\mathrm{p}}$, $P_{R_1}$, and $P_{R_2}$ are the losses in the plate capacitor $C_\mathrm{p}$ and the two resistors $R_1$ and $R_2$, respectively. This efficiency is higher than 90\% (maximum 94\% in the experiment) for a broad range of load resistances, as shown in Fig.~\ref{fig7-expeff}. Therefore, further increase of the overall efficiency can be realized by reducing the parasitic losses in  the operational amplifier, such as choosing alternative low-loss (and high-slew-rate) operational amplifiers or designing the amplifier topology using other efficient switching devices. In addition, the efficiency $\eta'$ (excluding the losses inside the operational amplifier) for the conventional WPT system [Fig.~4(a) with $\Rf=100~\mathrm{k\Omega}$]  is also calculated with LTspice, see Fig.~7. It is clear that the proposed WPT system indeed has higher efficiency than the conventional one as there is additional loss in $\Rf$.

\section{Conclusion}\label{sec:conclusion}

The proposed pulsed on-site wireless-power-generation paradigm based on the self-adaptive concept realizes a virtual, nearly ideal voltage source at the position of the power-receiving load. Under certain conditions, this voltage source is comparable to the ideal wired-connection case, enabling efficient and robust wireless power delivery. As the load and the capacitive wireless link are parts of the feedback loop of a self-oscillating circuit, the proposed WPT system automatically adjusts itself to the optimal working condition and provides robust operation against changes of the receiving end, including the load resistance and the receiver position. No additional tuning circuit is required to maintain the system in resonance which is needed in conventional WPT systems. While we demonstrate the robust WPT operation of the proposed system with a particular example, future work can be focused on the design of better circuit topologies, increase of the amplifier efficiency and the power level, and reduction of the apparatus size. In addition, safety concerns may arise for certain applications due to the use of electric field coupling. As such, the proposed design may be best suited for nonhuman environments (e.g., in automated factories). The proposed pulsed self-oscillation scheme can also be applied to inductive WPT systems, where the only requirement would be to co-design the whole system such that the coupling link and receiver are parts of the feedback loop, and which will be different from the solutions of adding additional sensing and control circuits~\cite{Namadmalan2016,Moghaddami2017}. On the basis of the theoretical analysis presented here, it is clear that the efficiency, load range, and transfer distance can be substantially increased with better-characteristic operational amplifiers or with use of other active switching components. In addition, this concept can potentially be used for wireless charging of smartphones, laptops, wearable devices, and electric vehicles, and can also be used to realize smart tables for multiple wireless-charging devices~\cite{Song2019,Liu2019arXiv}. The pulsed on-site WPT paradigm proposed and tested here has generally no fundamental limitations, and realizations with inductive coupling will also work.


\section*{Acknowledgments}
We thank Matti Vaaja for technical help with the experimental equipment, Viktar Asadchy for making the schematics, and Mohammad Sajjad Mirmoosa for useful discussions. 
This work was supported by the European Union's Horizon 2020 Future Emerging Technologies call (FETOPEN-RIA) under Grant Agreement No.~736876 (project
VISORSURF).

\bibliography{robustCWPT}

\begin{thebibliography}{10}
\bibitem{buffer}
The voltages $V_2$ and $V_3$ are measured through a buffer made from the same
  op-amp.
\end{thebibliography}
\bibliographystyle{bibstyle_science}

\end{document}